\documentclass[aps,prb,twocolumn,groupedaddress,showpacs]{revtex4}

\usepackage[dvips]{graphicx}
\usepackage{amsmath}
\usepackage{amstext}
\usepackage{graphics}
\usepackage{exscale}
\usepackage{epsfig}
\usepackage[T1]{fontenc}
\usepackage{bm}
\usepackage{bbm}

\usepackage{latexsym}

\renewcommand{\epsilon}{\varepsilon}

\newcommand{\Imag}{\mathrm{Im}}
\newcommand{\Real}{\mathrm{Re}}

\begin{document}

\title{Field dependent quasiparticles in a strongly correlated local system II}
\author{J. Bauer and A.C. Hewson}
\affiliation{Department of Mathematics, Imperial College, London SW7 2AZ,
  United Kingdom}
\date{\today} 
\begin{abstract}
We extend the renormalized quasiparticle  description  of the symmetric
 Anderson model in a magnetic field $H$, developed in earlier work,  
 to the non-symmetric model. The renormalized
parameters  are deduced from the low energy NRG fixed
point for arbitrary field values.
 We find 
quasiparticle resonance widths,   $\tilde\Delta_\sigma(H)$, 
which   depend in general on the spin $\sigma$ as well as $H$.
The low temperature static properties can be expressed completely in terms
of these parameters, which can also be used as inputs for a
renormalized
perturbation theory. We show that taking into account repeated
quasiparticle
scattering gives results for the
 longitudinal and transverse
dynamic spin susceptibilities which are in very good agreement with those
obtained
from  direct NRG calculations. 
 \end{abstract}

\maketitle

\section{Introduction}

In an earlier paper \cite{HBK06} (hereafter referred to as I), we  showed how the low energy
behavior of the 
 particle-hole  symmetric  Anderson impurity 
model\cite{And61} 
can be described in terms of  field-dependent quasiparticles. This model is
characterized by the three independent parameters, $\epsilon_d$, the impurity
level, $\Delta$, the broadening of this level due to the hybridization with
conduction electrons, and $U$, the interaction at the impurity site. In the
absence of a magnetic field, it was shown earlier \cite{HOM04,Hew93b,Hew05}
that the low energy behavior
can be 
described by an effective version of the {\em same } model with three corresponding
renormalized parameters, $\tilde\epsilon_d$,  $\tilde\Delta$, and $\tilde U$.
Subsequently, this approach was extended to include a  magnetic 
field $H$, and the  parameters were then found to be  field-dependent,
$\tilde\epsilon_{d,\sigma}(H)$,  $\tilde\Delta_\sigma(H)$, and $\tilde
U(H)$. One way to calculate these parameters  is from the low energy 
excitations of the numerical renormalization group (NRG) fixed point which was
used in the earlier paper I.  
A strong magnetic field tends to freeze the spin fluctuations and leads to a
de-renormalization of the quasiparticles. 
On increasing the field from zero in the strong coupling case the parameters
for the quasiparticles slowly revert to their uncorrelated mean field values
in the extreme high field limit.  

The renormalized parameters are not just a convenient way of describing the
low energy behavior; they  completely specify the model. A
renormalized perturbation theory (RPT) can be set up in which the free
propagators correspond to fully dressed quasiparticles
\cite{Hew93,Hew01}. This formalism is particularly effective for describing
the Fermi liquid regime, as only  diagrams up to second order have to be taken
into account to obtain asymptotically exact results for the $T=0$
susceptibilities, and the leading $T^2$ term in the conductivity (I). This
perturbation expansion is not restricted to the low energy and low 
temperature regime, and can be used for calculations on all energy scales. We
have  shown that a very good description of the $T=0$ spin and charge dynamics
for the Anderson model in the Kondo regime can be obtained  by summing the RPT
diagrams for repeated quasiparticle scattering \cite{Hew06}. The results give
an accurate description of the spin and charge susceptibilities for arbitrary
magnetic field values $H$, and for frequencies $\omega$ extending over a range
significantly larger than the Kondo temperature $T_{\rm K}$. The
Korringa-Shiba relation \cite{Shi75} and the sum rules for the spectral
density are satisfied. \par 

In this paper we show that this approach can be extended to the
non-symmetric Anderson model. There are significant differences in this case,
as the parameters acquire a spin dependence, and formulae given
earlier have to be generalized. 

\section{The non-symmetric Anderson model in a magnetic field}
The Hamiltonian for the Anderson model \cite{And61} is
\begin{eqnarray}
&&H_{\rm AM}=\sum\sb {\sigma}\epsilon\sb {\mathrm{d},\sigma} d\sp {\dagger}\sb
{\sigma}  
d\sp {}\sb {\sigma}+Un\sb {\mathrm{d},\uparrow}n\sb {\mathrm{d},\downarrow} \label{ham}\\
&& +\sum\sb {{ k},\sigma}( V\sb { k,\sigma}d\sp {\dagger}\sb {\sigma}
c\sp {}\sb {{ k},\sigma}+ V\sb { k,\sigma}\sp *c\sp {\dagger}\sb {{
k},\sigma}d\sp {}\sb {\sigma})+\sum\sb {{
k},\sigma}\epsilon\sb {{ k},\sigma}c\sp {\dagger}\sb {{ k},\sigma}
c\sp {}\sb {{
k},\sigma}, \nonumber
\end{eqnarray}
where $\epsilon_{\mathrm{d},\sigma}=\epsilon_{\rm d}-\sigma g\mu_{\rm B} H/2$
is the energy of the localized  level at an impurity site   in a magnetic
field $H$, $U$ the interaction at this local site, and $V_{k,\sigma}$ the
hybridization matrix element to a band of conduction electrons of spin
$\sigma$ with energy $\epsilon_{k,\sigma}-\sigma g_c\mu_{\rm B} H/2$, where
$g_c$ is the g-factor for the conduction electrons. When $U=0$ the local level
broadens into a resonance, corresponding to a localized quasi-bound state,
whose width depends on the quantity $ \Delta_\sigma(\omega)=\pi\sum\sb {k}|
V\sb {k,\sigma}|\sp 2\delta(\omega -\epsilon\sb { k,\sigma})$. For the
impurity model, where we are interested in universal features, it is usual to
take   a wide conduction band with a flat density of states so that
$\Delta_\sigma(\omega)$ becomes independent of $\omega$,
and can be taken as a constant $\Delta_\sigma$. In this wide band
limit $\Delta_\sigma(\omega)$ will be independent of the magnetic field
on the conduction electrons, so we can effectively put $g_c=0$.
When this is the case $\Delta_\sigma$ is usually taken to be a constant 
$\Delta$ independent of $\sigma$. 

\par
In the renormalized perturbation theory approach\cite{Hew93,Hew01} we cast the corresponding
Lagrangian for this model ${\cal L}_{\rm
AM}(\epsilon_{\mathrm{d},\sigma},\Delta,U)$  into  the form,
\begin{equation}
{\cal L}_{\rm AM}(\epsilon_{\mathrm{d},\sigma},
\Delta,U)={ \cal L}_{\rm AM}(\tilde\epsilon_{\mathrm{d},\sigma},
\tilde\Delta_\sigma,\tilde U)+ {\cal L}_{\rm
  ct}(\lambda_1,\lambda_2,\lambda_3),\label{lag}
\end{equation}
where the renormalized parameters,   $\tilde\epsilon_{\mathrm{d},\sigma}$ and 
$\tilde\Delta_{\sigma}$, are defined in terms of the self-energy
$\Sigma_{\sigma}(\omega)$ of the one-electron Green function for the impurity state,
\begin{equation}
G_{\sigma}(\omega)={1\over
    \omega-\epsilon_{\mathrm{d}\sigma}+i\Delta-\Sigma_\sigma(\omega)},
\label{gf}
\end{equation}
and are given by 
\begin{equation}
\tilde\epsilon_{\mathrm{d},{\sigma}}=z_{\sigma}(\epsilon_{\mathrm{d},{\sigma}} 
+\Sigma_\sigma(0)),\quad\tilde\Delta_{\sigma} =z_\sigma\Delta,
\label{ren1}
\end{equation} 
where $z_{\sigma}$ is given by
$z_{\sigma}={1/{(1-\Sigma_{\sigma}'(0))}}$.
The renormalized or quasiparticle interaction  $\tilde U$, is defined in terms
of the local total 4-vertex
$\Gamma_{\uparrow\downarrow}(\omega_1,\omega_2,\omega_3,\omega_4)$ at zero frequency,
  \begin{equation} 
\tilde U=z_{\uparrow}z_{\downarrow}\Gamma_{\uparrow\downarrow}(0,0,0,0).
\label{ren2}\end{equation}
It will be convenient to rewrite the spin dependent quasiparticle energies
in the form,
$\tilde\epsilon_{{\rm d},\sigma}=\tilde\epsilon_{\rm d}(h)-\sigma h\tilde\eta(h)$,
where
\begin{equation}
\tilde\epsilon_{\mathrm{d}}(h)={1\over
  2}\sum_\sigma\tilde\epsilon_{\mathrm{d},{\sigma}},\quad 
\tilde\eta(h)={1\over
  2h}\sum_\sigma \sigma\tilde\epsilon_{\mathrm{d},{\sigma}}
\end{equation}
where $\tilde\epsilon_{\rm d}(h)$ and $\tilde\eta(h)$ are both  even functions of the magnetic field
$h=g\mu_{\rm B}H/2$.\par
The renormalized perturbation expansion is in powers of the renormalized
interaction $\tilde U$ for the complete Lagrangian defined in equation
(\ref{lag}).  The counter term part of the Lagrangian $ {\cal L}_{\rm
  ct}(\lambda_1,\lambda_2,\lambda_3)$ essentially takes care of any 
overcounting.
\begin{figure}
\begin{center}
\includegraphics[width=0.45\textwidth,height=6cm]{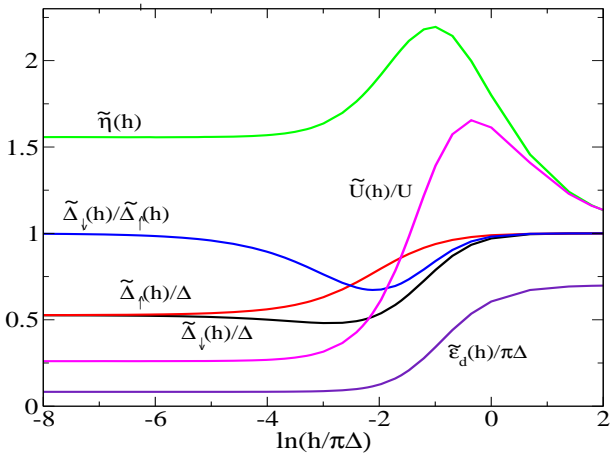}
\vspace*{-0.5cm}
\label{figure1}
\end{center} 
\caption{(Color online) Plots of the renormalized parameters,
  $\tilde\Delta_{\uparrow}(h)/\Delta$, $\tilde\Delta_{\downarrow}(h)/\Delta$,
 $\tilde\epsilon_{\rm d}(h)/\pi\Delta$, $\tilde U(h)/U$, $\tilde\eta(h)$, for
 the asymmetric Anderson model, with $\pi\Delta=0.1$, $U/\pi\Delta=2$ and
 $\epsilon_d/\pi\Delta=-0.3$, as a function of the logarithm of the magnetic
 field $h/\pi\Delta$. The ratio
 $\tilde\Delta_{\downarrow}(h)/\tilde\Delta_{\uparrow}(h)$ is also shown.}
\label{figure1}
\end{figure}
The parameters, $\tilde\epsilon_{\mathrm{d},{\sigma}}$, $\tilde\Delta_\sigma$
 and $\tilde U$, have been taken to be the fully renormalized ones.
The most effective way of estimating them in this case is from the energy
levels of a numerical renormalization group (NRG) calculation
 \cite{HOM04,Hew05}. However, the approach itself is independent of the NRG,
 and the parameters can be deduced in other ways, for instance from experiment.
The  counter term parameters, $\lambda_1$, $\lambda_2$ and $\lambda_3$, are 
required to cancel any further renormalization, and are completely determined
by this condition \cite{Hew93,Hew01}. 

In figure \ref{figure1} we display some typical results for  renormalized
parameters as a function of the magnetic 
field on a log scale. For this plot we have taken bare parameters $\epsilon_{\rm d}/\pi\Delta=-0.3$
and $U/\pi\Delta=2$, corresponding to a impurity occupation in the absence of
a field, $\langle
n_{d,\sigma}\rangle=n(0)/2\sim 0.35$. With these values the parameters are
strongly renormalized for $h=0$. 
The overall trend as a function of $h$ is very similar to
that for the particle-hole symmetric case (I) in the strong coupling regime, 
with the parameters converging to the mean field values in the limit $h\to
\infty$. We do see, however, that the resonance widths become spin dependent,
$\Delta_{\uparrow}(h)\ne\Delta_{\downarrow}(h)$, except asymptotically as
$h\to 0$ and $h\to\infty$. We note that, though 
$\tilde\Delta_{\uparrow}(h)$ increases monotonically with increase of $h$,  
$\tilde\Delta_{\downarrow}(h)$ initially decreases. In this case, where the
impurity level is less than half-filled, the ratio
$\tilde\Delta_{\uparrow}(h)/\tilde\Delta_{\downarrow}(h)\ge 1$. This ratio is
reversed, so $\tilde\Delta_{\uparrow}(h)/\tilde\Delta_{\downarrow}(h)\le 1$,
when the impurity level is more than half-filled. 

\section{Static response functions}
Once the renormalized parameters have been determined, the static impurity response
functions can be evaluated directly by substituting into the relevant {\it
  exact} formula. This is a more direct way of calculating these response
functions than the usual NRG way, which involves a subtraction procedure to
estimate the impurity component \cite{Wil75,KWW80a,KWW80b}. 
The formulae in the absence of a magnetic field
\cite{HOM04}, and for the symmetric model in the presence of a field, were
given earlier (I). Here, we give the generalizations for the
non-symmetric model. The induced magnetization $M(h)$ is given by
$M(h)=g\mu_{\rm B}m(h)$, where 
\begin{equation}
m(h)={1\over 2}(n_{{\rm d}\uparrow}-n_{{\rm d}\downarrow})={- 1\over 2\pi}\sum_\sigma
\sigma\,{\rm tan}^{-1}
\left(\frac{\tilde\epsilon_{{\rm d}\sigma}(h)}{\tilde\Delta_{\sigma}(h)}\right),
\label{magqp}
\end{equation}
which can be derived from the Friedel sum rule. The 
longitudinal  susceptibility  $\chi_l(h)$ (in units of $(g\mu_{\rm B})^2$)
is given by
\begin{equation}
\chi_l(h)=0.25(\tilde\rho_\uparrow(0,h)+\tilde\rho_\downarrow(0,h)+\tilde
U(h)\tilde\rho_\uparrow(0,h)\tilde\rho_\downarrow(0,h)),
\label{chil}
\end{equation}
where $\tilde\rho_\sigma(\omega,h)$ is the free quasiparticle density of states
given by 
\begin{equation}
\tilde\rho_\sigma(\omega,h)={1\over \pi}{\tilde\Delta_\sigma(h)\over
(\omega-\tilde\epsilon_{{\rm
    d},\sigma}(h))^2+\tilde\Delta^2_\sigma(h)}.
\label{qpdos}
\end{equation}
The corresponding transverse susceptibility $\chi_t(h)$ (zero applied field
limit in the transverse direction) is given by 
\begin{equation}
\chi_t(h)=\frac{m(h)}{2h}.
\label{chit}
\end{equation}
The total occupation of the impurity site $n(h)=(n_{{\rm d}\uparrow}+n_{{\rm
  d}\downarrow})$ can be derived similarly, and is given by
\begin{equation}
n(h)=1-{1\over 2\pi}\sum_\sigma {\rm tan}^{-1} 
\left(\frac{\tilde\epsilon_{{\rm
        d}{\sigma}}(h)}{\tilde\Delta_{\sigma}(h)}\right),
\label{nqp}
\end{equation}
and the local charge susceptibility  $\chi_c(h)$ is given by
\begin{equation}
\chi_c(h)=0.25[\tilde\rho_\uparrow(0,h)+\tilde\rho_\downarrow(0,h)-\tilde
U(h)\tilde\rho_\uparrow(0,h)\tilde\rho_\downarrow(0,h)] .
\label{chic}
\end{equation}

To illustrate the  magnetic field dependence of these static response functions,
we evaluate these formulae in a particular case using the field dependent
renormalized parameters given in figure \ref{figure1}. The results are compared with
those obtained by a evaluation of static expectation values using the
NRG. In all subsequent calculations in this paper, we concentrate on this same set
of bare parameters, $\epsilon_{\rm d}/\pi\Delta=-0.3$ and $U/\pi\Delta=2$.

Shown in figure \ref{figure2} is the result for  $m(h)$ deduced from equation 
(\ref{magqp}), compared with results obtained by the direct evaluation of the
d-site occupation values in the ground state as determined from the NRG. 
\begin{figure}
\begin{center}
\includegraphics[width=0.45\textwidth,height=6cm]{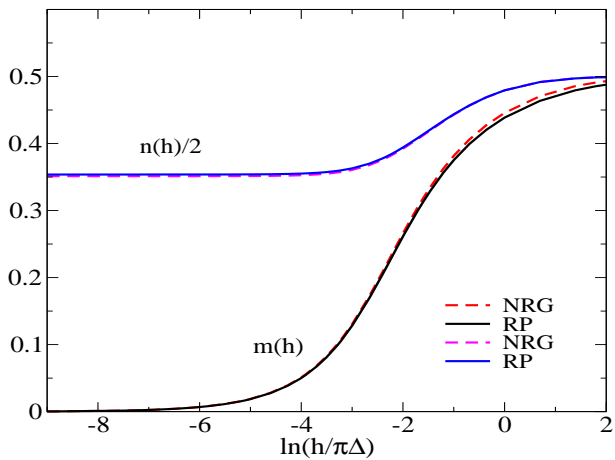}
\vspace*{-0.5cm}
\label{figure2}
\end{center} 
\caption{(Color online) The induced magnetization $ m(h)$ as a function of the logarithm of
  the  magnetic field $h$ for the asymmetric Anderson model with the same set
  of parameters as given in figure \ref{figure1}. The dashed curve is that calculated from
  the direct evaluation of the occupation values from the NRG ground state,
  and the full curve is that deduced from the renormalized parameters in
  equation (\ref{magqp}). Also shown is the average occupation $n(h)/2$ as
  calculated from the NRG ground state (dashed curve) and the quasiparticle
  occupation values (full curve) as given in equation (\ref{nqp}).}
\label{figure2}
\end{figure}
There is a small but systematic difference, of the order of 2\%,  between the
two sets of results. This difference could be due to the fact that
we assume an infinite bandwidth in the derivation of these formulae, whereas
we take a finite band width $2D$, with $D=1$ in the NRG calculations. 
The corresponding estimates of the  average occupation
number $n(h)/2$  as a function of magnetic field $h$ are shown in the same
figure. For this quantity the two sets of results are almost indistinguishable.
In the extreme large field limit the average occupation of the impurity
level tends to unity, as the majority spin level gets pulled further and
further below the Fermi level, and the impurity becomes completely polarized.
In this regime the average renormalized level, $\tilde\epsilon_{\rm d}$,
approaches the mean field value, 
$\tilde\epsilon_{\rm d}=\epsilon_{\rm d}+0.5Un(h)$, rather than the bare
value $\epsilon_{\rm d}$, while the other renormalized quantities approach
their bare values.\par
The longitudinal and transverse spin susceptibilities,  $\chi_l(h)$ and
$\chi_t(h)$,  are plotted in figure \ref{figure3} as a function of the
 logarithm of  the  magnetic field $h$.   
\begin{figure}
\begin{center}
\includegraphics[width=0.45\textwidth,height=6cm]{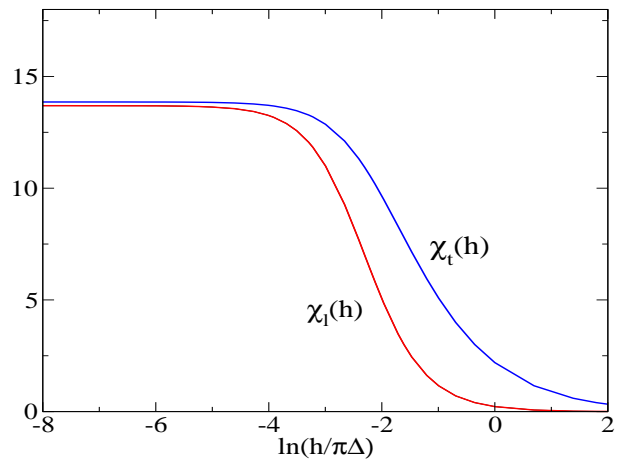}
\vspace*{-0.5cm}
\label{figure3}
\end{center} 
\caption{(Color online) The longitudinal and transverse impurity site susceptibilities,
  $\chi_l(h)$ and  $\chi_t(h)$,  as a function of the logarithm of  the
 magnetic field $h$.  $\chi_l(h)$ is calculated from equation
 (\ref{chil}) and  $\chi_t(h)=m(h)/2h$. }
\vspace{0.5cm}
\label{figure3}
\end{figure}
$\chi_l(h)$ is calculated from equation (\ref{chil}) and  $\chi_t(h)=m(h)/2h$ 
as calculated from equation (\ref{magqp}). 
They should asymptotically converge to the same result in the limit $h\to 0$.
There seems to be a very small discrepancy, of about 1\%, between the two estimates in this
limit. This could be due to evaluating the ratio in equation (\ref{chit}) for
very small values of the field.

We now look at the behavior of the corresponding dynamic response functions.

\section{Dynamic response functions}

\subsection{Single particle spectra}

It is of interest to see how well the free quasiparticle density of states 
given in equation (\ref{qpdos}) compares with the spectral density
$\rho_\sigma(\omega)$ calculated from the NRG for each spin type in the
presence of a field. 
The NRG spectra are calculated from the self-energy following the
prescription given in \cite{BHP98}, and we have also used the improved method
\cite{PPA06,WD06pre} based on the complete Anders-Schiller basis \cite{AS05}.

In figure \ref{figure4} (upper panel) we show $\rho_\uparrow(\omega)$  and
$\rho_\downarrow(\omega)$, as calculated from the NRG, for the field
$h/\pi\Delta=0.001$.

\begin{figure}
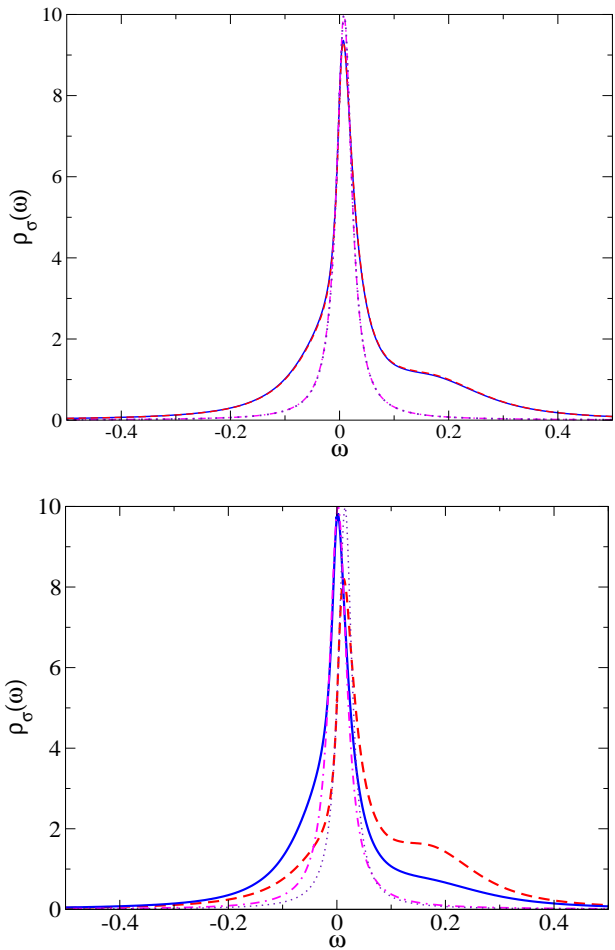

\begin{center}
\includegraphics[width=0.45\textwidth,height=6cm]{figures_alex/rho_0.0001.eps}\\[.5cm]
\includegraphics[width=0.45\textwidth,height=6cm]{figures_alex/rhoh_0.004.eps}
\vspace*{-0.5cm}
\label{figure4}
\end{center} 
\caption{(Color online) The spectral density for the spin up (majority) electrons
  $\rho_{\uparrow}(\omega,h)$  (full curve) and spin down (minority) electrons
  $\rho_{\downarrow}(\omega,h)$ (dashed curve) at $T=0$ as a function of
  $\omega$ for  $h/\pi\Delta=0.001$ (upper panel) and $h/\pi\Delta=0.04$
  (lower panel). The two curves on the upper panel are almost coincident on
  the scale shown. Also shown are the corresponding quantities derived from
  the free quasiparticle densities of states,
  $z_\uparrow\tilde\rho_{\uparrow}(\omega)$ (dot-dashed) and
  $z_\downarrow\tilde\rho_{\downarrow}(\omega)$ (dotted).}
\label{figure4}
\end{figure}
\noindent
We see that in this relatively weak field that there are  only a small differences between the two spectral
densities; they are almost coincident on the scale shown. 
The corresponding quantities derived from the free quasiparticle densities of
states, $z_\uparrow\tilde\rho_{\uparrow}(\omega)$ and
$z_\downarrow\tilde\rho_{\downarrow}(\omega)$, are shown in the same figure.

In figure \ref{figure4} (lower panel) we make a similar comparison for a large
field $h/\pi\Delta=0.04$, and the same set of bare parameters. The
polarization is much stronger for this higher magnetic field value and there
is now a marked difference between the two spectral densities for the two spin
types. The spectra derived from the free quasiparticles, can be  seen to
describe the spectrum in the immediate vicinity of the Fermi level for both
spin types. They cannot describe the difference in the heights of the peaks of
the two spin types, because in these simple Lorentzian formula the height of
peaks in the quasiparticle weighted spectra are  independent of $H$ and the
spin and given by $1/\pi\Delta$.
Note that the description of the low frequency range can be extended within
the renormalized perturbation theory frame work by including a self-energy
correction in the quasiparticle density of states as shown in the particle
hole symmetric case \cite{BHO07}. 

In the next section we look beyond the simple free quasiparticle picture and
take account of the leading quasiparticle scattering terms. 


\subsection{Dynamic spin susceptibilities}

\begin{figure}
\begin{center}
\includegraphics[width=0.4\textwidth,height=3.3cm]{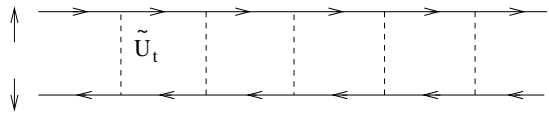}
\vspace*{-2.5cm}
\end{center} 
\caption{Repeated scattering of a quasiparticle with spin $\uparrow$ and
a quasihole with spin $\downarrow$ via the effective interaction $\tilde U_t$. }
\label{figure6}
\end{figure}

The leading corrections to the two particle dynamic response functions to the
free quasiparticle picture in the renormalized perturbation theory arise from
the diagrams with repeated quasiparticle scattering. In figure \ref{figure6}
we illustrate this type of diagram for the scattering of a free up spin
quasiparticle with a free down spin quasihole. 

\noindent
Such a diagram contributes to the transverse dynamic susceptibility $\chi_{t}(\omega,h)$ in the
presence of the magnetic field. We derived a quasiparticle interaction
term $\tilde U(h)$ earlier, but this is not the interaction term we need in
considering this diagram, as this corresponds to the total 4-vertex, 
$\tilde\Gamma_{\uparrow\downarrow}(0,0,0,0)$,
which has implicitly included this scattering for $\omega=0$. What we
use is the irreducible particle-hole interaction at zero frequency in this channel,
which we denote by $\tilde U_t(h)$. The dynamic transverse susceptibility
$\chi_{t}(\omega,h)$, is then  given by 
\begin{equation}
\chi_{t}(\omega,h)={\tilde\chi_{\uparrow\downarrow}(\omega,h)
\over 1-\tilde
U_t(h)\tilde\chi_{\uparrow\downarrow}(\omega,h)},
\label{trrpt}
\end{equation}
where $\tilde\chi_{\uparrow\downarrow}(\omega,h)$ is the transverse spin
susceptibility of the free quasiparticle. The analytic expression for 
 $\tilde\chi_{\uparrow\downarrow}(\omega,h)$ is given in the appendix.\par
To find the unknown quantity  $\tilde U_t(h)$ we use the fact that this series
for $\omega=0$ gives the static transverse susceptibility which
we have calculated already, and is given in equation (\ref{chit}). Hence,
comparing (\ref{trrpt}) for $\omega=0$ with equation (\ref{chit}) yields
\begin{equation}
\tilde U_t(h)={1\over m(h)}
\left\{{4\tilde h^2+(\tilde\Delta_\uparrow-\tilde\Delta_\downarrow)^2
\over 4\tilde h+{(\tilde\Delta_\uparrow-\tilde\Delta_\downarrow)
\over 2\pi m(h)}\,{\rm ln}\left({(\tilde
  h-\tilde\epsilon_d)^2+\tilde\Delta_\uparrow^2\over (\tilde
  h+\tilde\epsilon_d)^2+\tilde\Delta_\downarrow^2}\right)}-h\right\},
\end{equation}
where $\tilde h=h\tilde\eta(h)$, and  $m(h)$ is given in equation
(\ref{magqp}). This expression simplifies in the case,
$\tilde\Delta_\uparrow=\tilde\Delta_\downarrow$, to give $\tilde
U_t(h)=(\tilde h -h)/m(h)$, which is the same as that used
earlier \cite{Hew06}.

Now we test how well these formulae describe the spin dynamic response over
the relevant frequency range. To do this we compare the predictions based on
the RPT formula (\ref{trrpt}) with those obtained from a direct NRG
evaluation.

\begin{figure}
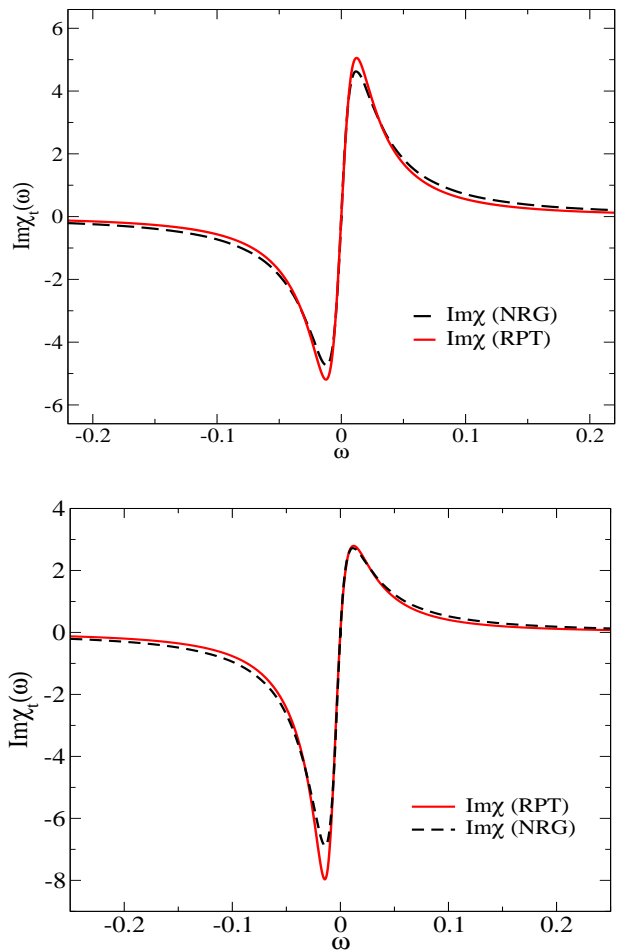

\begin{center}
\includegraphics[width=0.45\textwidth,height=6cm]{figures_alex/nchit_h_0.0001.eps}\\[0.5cm]
\includegraphics[width=0.45\textwidth,height=6cm]{figures_alex/tchit_h_0.004.eps}
\vspace*{-0.5cm}
\end{center} 
\caption{(Color online) The imaginary part of the transverse dynamic susceptibility,
  $\chi_t(\omega,h)$  at $T=0$ as a function of $\omega$, 
 for $h/\pi\Delta=0.001$ (upper panel) and $h/\pi\Delta=0.04$ (lower
 panel). The dashed curve is calculated from a direct NRG 
  calculation and the full curve from equation (\ref{trrpt}) with the
  renormalized parameters. }
\label{figure7}
\end{figure}

The imaginary part of the transverse dynamic susceptibility is shown in
figure \ref{figure7} (upper panel) for a magnetic field value $h/\pi\Delta=0.001$. 
The dashed curve is that from a direct NRG evaluation \cite{SSK89,CHZ94} and the full curve is that
calculated using equation (\ref{trrpt}), with the corresponding renormalized parameters.
There is very good agreement between the two sets of results.

Results for the imaginary part of $\chi_{t}(\omega,h)$  for the field value
$h/\pi\Delta=0.04$ are shown in figure \ref{figure7} (lower panel). In this
stronger field one of the peaks is suppressed while the other peak is
enhanced. The peak positions are in good agreement, and the slightly broader
peak  from the NRG data can be attributed to the logarithmic broadening 
used in the direct NRG evaluation.  There is a sum rule, that the total
spectral weight is equal to $-2m(h)$, which is satisfied precisely both in
the RPT result and also in the NRG calculation, as we have used the improved
prescription for the response functions based on the complete
Anders-Schiller basis.

In a similar way, we can calculate the longitudinal dynamic susceptibility
$\chi_{l}(\omega,h)$ from repeated scattering of the spin up and spin down
quasiparticles.  The irreducible zero frequency interaction in this channel we
denote by $\tilde U_l(h)$, and determine it from the condition,
$\chi_l(0,h)=\chi_l(h)$, using equation ({\ref{chil}).  The results for
$\chi_{l}(\omega,h)$ is 
\begin{eqnarray}
&&\chi_{l}(\omega,h)=  \label{lrrpt} \\
&&{\tilde\chi_{\uparrow\uparrow}(\omega,h)
 +\tilde\chi_{\downarrow\downarrow}(\omega,h)+4\tilde
 U_l(h)\tilde\chi_{\uparrow\uparrow}(\omega,h)\tilde\chi_{\downarrow\downarrow}(\omega,h)\over 2(
 1-4\tilde
 U^2_l(h)\tilde\chi_{\uparrow\uparrow}(\omega,h)\tilde\chi_{\downarrow\downarrow}(\omega,h))},
\nonumber
 \end{eqnarray}
where the analytic expression for $\tilde\chi_{\sigma\sigma}(\omega,h)$
is given in the Appendix. Equation (\ref{lrrpt}) is a generalization to  our
earlier result \cite{Hew06} for the particle-hole symmetric model. The zero
frequency irreducible particle-hole vertex $\tilde U_l(h)$ in this scattering
channel is given by  
\begin{equation}
\tilde U_l(h)=
\frac{-1+\sqrt{[1+\tilde
    U^2(h)(\tilde\rho_\uparrow+\tilde\rho_\downarrow)]^2-\tilde
    U^2(h)(\tilde\rho_\uparrow-\tilde\rho_\downarrow)^2}}
  {2(\tilde\rho_\uparrow+\tilde\rho_\downarrow+\tilde U(h)
\tilde\rho_\uparrow\tilde\rho_\downarrow)},
\end{equation}
where we have simplified the notation,
$\tilde\rho_\sigma(0,h)=\tilde\rho_\sigma$. In the absence of a magnetic
field, or  with a magnetic field for the particle-hole symmetric model, 
    $\tilde\rho_\uparrow=\tilde\rho_\downarrow$, the result
simplifies to $\tilde U_l(h)=\tilde U(h)/[1+\tilde U(h)\tilde\rho(0,h)]$,
which is the value used in the earlier work\cite{Hew06}.


The imaginary part of the longitudinal dynamic spin
susceptibility in the weak field case, $h/\pi\Delta=0.001$, 
is similar to that for the transverse case shown in figure \ref{figure7}
(upper panel), apart from an overall factor of 2, so we do not show he results
here. The stronger field case,  $h/\pi\Delta=0.04$, is shown in figure 
\ref{figure9}. 

\begin{figure}
\begin{center}
\includegraphics[width=0.45\textwidth,height=6cm]{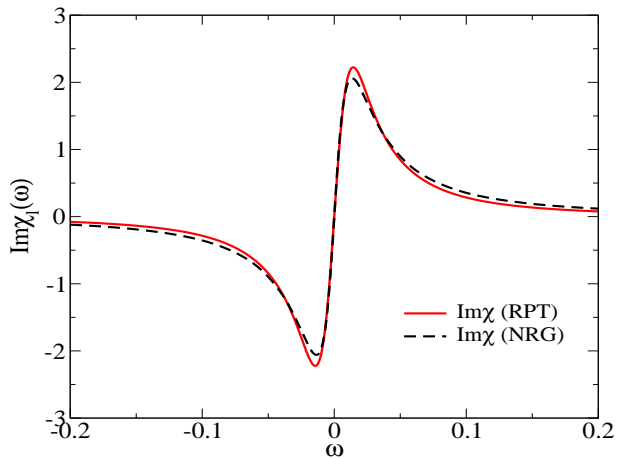}
\vspace*{-0.5cm}
\end{center} 
\caption{(Color online) The imaginary part of the longitudinal dynamic susceptibility,
  $\chi_l(\omega,h)$  at $T=0$ as a function of $\omega$  for
  ($h/\pi\Delta=0.04$).}
\label{figure9}
\end{figure}
\noindent
Due to the stronger magnetic field, the heights of the peaks are slightly
reduced as compared with the results  for $h/\pi\Delta=0.001$, but the
overall features are very similar.  The Korringa-Shiba relation does not hold
for the model without particle-hole symmetry in the presence of a magnetic
field. Where it  does hold, in the absence of a field, or with particle-hole
symmetry, the renormalized perturbation expression satisfies it exactly\cite{Hew06}.  

\section{Summary and Discussion} 
In this paper we have extended our earlier work I, where we described  the low
energy  behavior  of the symmetric  Anderson model in a magnetic field $h$ in
terms of  field-dependent renormalized quasiparticles, to the non-symmetric 
Anderson model. The main new feature that emerges is the dependence of
the quasiparticle peak resonance width $\tilde\Delta_\sigma(h)$ 
on the spin type $\sigma$ as well as on the value of the magnetic field.
The $T=0$ spin and charge susceptibilities can be expressed as
exact formulae in terms of the parameters $\Delta_\sigma(h)$,
$\tilde\epsilon_{{\rm d},\sigma}(h)$ and a local field dependent interaction
between the quasiparticles $\tilde U(h)$. 
Based on these parameters the low temperature behavior of the model, such as
the susceptibilities and conductance, in the presence of the field could be
calculated as described for the symmetric model in I. 
It was also demonstrated earlier that an excellent description of the low
energy spin dynamics can be  obtained for the symmetric model \cite{Hew06} 
based on approximate formulae which take into account repeated quasiparticle
scattering in the RPT. Here these results have been generalized to the
non-symmetric model for the transverse and longitudinal spin susceptibilities,
which again agree remarkably well with those obtained from a direct NRG
calculation.  

What may seem surprising at first is that the results for the dynamic
susceptibilities,  based on the repeated quasiparticle scattering, agree so
well over the full range of $\omega$, whereas the free quasiparticle density
of states only describes the single electron spectral densities well only in
the immediate region of the 
Fermi level. However, as discussed in I, the quasiparticle density of states
$\tilde\rho_\sigma(\omega)$, survives in the limit $z_\sigma\to 0$, when it
becomes a delta function describing a free spin. For very small $z_\sigma$
the product $z_\sigma\tilde\rho_\sigma(\omega)$, makes very little
contribution to the spectral density $\rho_\sigma(\omega)$, except near
the Fermi level. Hence in the strong correlation regime, when $z_\sigma$ is
very small, it would appear to be more appropriate to interpret the quasiparticle
as describing a spinon excitation. Because the dominant low
energy excitations are spinons, as is known from the Bethe ansatz solutions 
for the Kondo model \cite{TW83,AFL83}, then it is not so surprising that they
provide a good description of the spin dynamics.

\vspace*{1cm}
\noindent{\bf Acknowledgement}

\bigskip
\noindent
We wish to thank N. Dupuis, D.M. Edwards, W. Koller, D. Meyer and A. Oguri for helpful
discussions and   W. Koller and D. Meyer for their contributions to the
development of the NRG programs. 
One of us (J.B.) thanks the Gottlieb Daimler and Karl Benz Foundation, the
German Academic exchange service (DAAD) and the EPSRC for financial support.

\bigskip
\section{Appendix} 
The free quasiparticle dynamic susceptibility
$\tilde\chi_{\sigma,\sigma'}(\omega)$ for the impurity model in the wide band
limit, $\tilde\Delta_{\uparrow}=\tilde\Delta_{\downarrow}$, were given earlier
\cite{Hew06}. Here we give the more general results for
$\tilde\Delta_{\uparrow}\neq\tilde\Delta_{\downarrow}$,
\begin{equation}
\tilde\chi_{\sigma,\sigma}(\omega)
={-1\over \pi\omega}
{\tilde\Delta_\sigma\over \omega-2i\tilde\Delta_\sigma}\sum_{\alpha=-1,1}{\rm
    ln}\left(1-{\omega\over{\alpha\tilde\epsilon_{d,\sigma}+i\tilde\Delta_\sigma}}\right),
\label{chizz}
\end{equation}
for $\omega>0$, and for $\omega=0$,
\begin{equation}
\tilde\chi_{\sigma,\sigma}(0)
=\tilde\rho_{\sigma}(0).
\end{equation}
The values for $\omega<0$ follow from the fact that
$\Real\tilde\chi_{\sigma,\sigma}(\omega)=\Real\tilde\chi_{\sigma,\sigma}(-\omega)$ and
$\Imag\tilde\chi_{\sigma,\sigma}(\omega)=-\Imag\tilde\chi_{\sigma,\sigma}(-\omega)$.
For $\sigma'\neq\sigma$, 
\begin{eqnarray*}
&&\tilde\chi_{\uparrow,\downarrow}(\omega)= \\
&&\frac{i/2\pi}
{( \omega+\tilde\epsilon_{d,\downarrow}-\tilde\epsilon_{d,\uparrow}
  +i\tilde\Delta_\uparrow-i\tilde\Delta_\downarrow)}
{\rm  ln}\left({\omega-\tilde\epsilon_{d,\uparrow}-i\tilde\Delta_\uparrow\over
 -i\tilde\Delta_\downarrow-\tilde\epsilon_{d,\downarrow}}\right) \\
&&+{i/2\pi\over (\omega+
  \tilde\epsilon_{d,\downarrow}-\tilde\epsilon_{d,\uparrow}-i\tilde\Delta_\uparrow+i
\tilde\Delta_\downarrow)}{\rm
ln}\left({\omega+\tilde\epsilon_{d,\downarrow}-i\tilde\Delta_\downarrow\over   
 -i\tilde\Delta_\uparrow+\tilde\epsilon_{d,\uparrow}}\right) \\
&&+ {-i/2\pi\over (\omega+
  \tilde\epsilon_{d,\downarrow}-\tilde\epsilon_{d,\uparrow}+i\tilde\Delta_\uparrow+i\tilde\Delta_\downarrow)} \times \\  
&&\times \left[{\rm ln}\left(1+{\omega\over
      i\tilde\Delta_\uparrow-\tilde\epsilon_{d,\uparrow}}\right) +{\rm
    ln}\left(1+{\omega\over
      i\tilde\Delta_\downarrow+\tilde\epsilon_{d,\downarrow}}\right)\right] .
\end{eqnarray*}

\bibliography{artikel,biblio1}

\end{document}